
\catcode`\@=11
\font\tensmc=cmcsc10      
\def\smc{\tensmc}

\def\hcorrection#1{\advance\hoffset by #1 }
\def\vcorrection#1{\advance\voffset by #1 }
\def\wlog#1{}
\newif\iftitle@
\outer\def\title{\title@true\vglue 24\p@ plus 12\p@ minus 12\p@
   \bgroup\let\\=\cr\tabskip\centering
   \halign to \hsize\bgroup\tenbf\hfill\ignorespaces##\unskip\hfill\cr}
\def\endtitle{\cr\egroup\egroup\vglue 18\p@ plus 12\p@ minus 6\p@}
\outer\def\author{\iftitle@\vglue -18\p@ plus -12\p@ minus -6\p@\fi\vglue
    12\p@ plus 6\p@ minus 3\p@\bgroup\let\\=\cr\tabskip\centering
    \halign to \hsize\bgroup\smc\hfill\ignorespaces##\unskip\hfill\cr}
\def\endauthor{\cr\egroup\egroup\vglue 18\p@ plus 12\p@ minus 6\p@}
\outer\def\heading{\bigbreak\bgroup\let\\=\cr\tabskip\centering
    \halign to \hsize\bgroup\smc\hfill\ignorespaces##\unskip\hfill\cr}
\def\endheading{\cr\egroup\egroup\nobreak\medskip}

\outer\def\endproclaim{\par\ifdim\lastskip<\medskipamount\removelastskip
  \penalty 55 \fi\medskip\rm}
\outer\def\demo#1{\par\ifdim\lastskip<\smallskipamount\removelastskip
    \smallskip\fi\noindent{\smc\ignorespaces#1\unskip:\enspace}\rm
      \ignorespaces}

\newcount\footmarkcount@
\footmarkcount@=1
\def\makefootnote@#1#2{\insert\footins{\interlinepenalty=100
  \splittopskip=\ht\strutbox \splitmaxdepth=\dp\strutbox
  \floatingpenalty=\@MM
  \leftskip=\z@\rightskip=\z@\spaceskip=\z@\xspaceskip=\z@
  \noindent{#1}\footstrut\rm\ignorespaces #2\strut}}
\def\footnote{\let\@sf=\empty\ifhmode\edef\@sf{\spacefactor
   =\the\spacefactor}\/\fi\futurelet\next\footnote@}
\def\footnote@{\ifx"\next\let\next\footnote@@\else
    \let\next\footnote@@@\fi\next}
\def\footnote@@"#1"#2{#1\@sf\relax\makefootnote@{#1}{#2}}
\def\footnote@@@#1{$^{\number\footmarkcount@}$\makefootnote@
   {$^{\number\footmarkcount@}$}{#1}\global\advance\footmarkcount@ by 1 }

\hyphenation{man-u-script man-u-scripts ap-pen-dix ap-pen-di-ces}
\hyphenation{data-base data-bases}
\ifx\amstexloaded@\relax\catcode`\@=13
  \endinput\else\let\amstexloaded@=\relax\fi
\newlinechar=`\^^J
\def\eat@#1{}
\def\Space@.{\futurelet\Space@\relax}
\Space@. %
\newhelp\athelp@
{Only certain combinations beginning with @ make sense to me.^^J
Perhaps you wanted \string\@\space for a printed @?^^J
I've ignored the character or group after @.}
\def\futureletnextat@{\futurelet\next\at@}
{\catcode`\@=\active
\lccode`\Z=`\@ \lowercase
{\gdef@{\expandafter\csname futureletnextatZ\endcsname}
\expandafter\gdef\csname atZ\endcsname
   {\ifcat\noexpand\next a\def\next{\csname atZZ\endcsname}\else
   \ifcat\noexpand\next0\def\next{\csname atZZ\endcsname}\else
    \def\next{\csname atZZZ\endcsname}\fi\fi\next}
\expandafter\gdef\csname atZZ\endcsname#1{\expandafter
   \ifx\csname #1Zat\endcsname\relax\def\next
     {\errhelp\expandafter=\csname athelpZ\endcsname
      \errmessage{Invalid use of \string@}}\else
       \def\next{\csname #1Zat\endcsname}\fi\next}
\expandafter\gdef\csname atZZZ\endcsname#1{\errhelp
    \expandafter=\csname athelpZ\endcsname
      \errmessage{Invalid use of \string@}}}}
\def\atdef@#1{\expandafter\def\csname #1@at\endcsname}
\newhelp\defahelp@{If you typed \string\define\space cs instead of
\string\define\string\cs\space^^J
I've substituted an inaccessible control sequence so that your^^J
definition will be completed without mixing me up too badly.^^J
If you typed \string\define{\string\cs} the inaccessible control sequence^^J
was defined to be \string\cs, and the rest of your^^J
definition appears as input.}
\newhelp\defbhelp@{I've ignored your definition, because it might^^J
conflict with other uses that are important to me.}
\def\define{\futurelet\next\define@}
\def\define@{\ifcat\noexpand\next\relax
  \def\next{\define@@}%
  \else\errhelp=\defahelp@
  \errmessage{\string\define\space must be followed by a control
     sequence}\def\next{\def\garbage@}\fi\next}
\def\undefined@{}
\def\preloaded@{}
\def\define@@#1{\ifx#1\relax\errhelp=\defbhelp@
   \errmessage{\string#1\space is already defined}\def\next{\def\garbage@}%
   \else\expandafter\ifx\csname\expandafter\eat@\string
         #1@\endcsname\undefined@\errhelp=\defbhelp@
   \errmessage{\string#1\space can't be defined}\def\next{\def\garbage@}%
   \else\expandafter\ifx\csname\expandafter\eat@\string#1\endcsname\relax
     \def\next{\def#1}\else\errhelp=\defbhelp@
     \errmessage{\string#1\space is already defined}\def\next{\def\garbage@}%
      \fi\fi\fi\next}
\def\famzero{\fam\z@}

\def\exp{\mathop{\famzero exp}\nolimits}

\def\lim{\mathop{\famzero lim}}

\def\textfont@#1#2{\def#1{\relax\ifmmode
    \errmessage{Use \string#1\space only in text}\else#2\fi}}
\textfont@\rm\tenrm
\textfont@\it\tenit
\textfont@\sl\tensl
\textfont@\bf\tenbf
\textfont@\smc\tensmc
\let\ic@=\/
\def\/{\unskip\ic@}
\def\textfonti{\the\textfont1 }
\def\t#1#2{{\edef\next{\the\font}\textfonti\accent"7F \next#1#2}}
\let\B=\=
\let\D=\.
\def~{\unskip\nobreak\ \ignorespaces}
{\catcode`\@=\active
\gdef\@{\char'100 }}
\atdef@-{\leavevmode\futurelet\next\athyph@}
\def\athyph@{\ifx\next-\let\next=\athyph@@
  \else\let\next=\athyph@@@\fi\next}
\def\athyph@@@{\hbox{-}}
\def\athyph@@#1{\futurelet\next\athyph@@@@}
\def\athyph@@@@{\if\next-\def\next##1{\hbox{---}}\else
    \def\next{\hbox{--}}\fi\next}
\def\.{.\spacefactor=\@m}
\atdef@.{\null.}
\atdef@,{\null,}
\atdef@;{\null;}
\atdef@:{\null:}
\atdef@?{\null?}
\atdef@!{\null!}
\def\srdr@{\thinspace}
\def\drsr@{\kern.02778em}
\def\sldl@{\kern.02778em}
\def\dlsl@{\thinspace}
\atdef@"{\unskip\futurelet\next\atqq@}
\def\atqq@{\ifx\next\Space@\def\next. {\atqq@@}\else
         \def\next.{\atqq@@}\fi\next.}
\def\atqq@@{\futurelet\next\atqq@@@}
\def\atqq@@@{\ifx\next`\def\next`{\atqql@}\else\def\next'{\atqqr@}\fi\next}
\def\atqql@{\futurelet\next\atqql@@}
\def\atqql@@{\ifx\next`\def\next`{\sldl@``}\else\def\next{\dlsl@`}\fi\next}
\def\atqqr@{\futurelet\next\atqqr@@}
\def\atqqr@@{\ifx\next'\def\next'{\srdr@''}\else\def\next{\drsr@'}\fi\next}

\def\textfontii{\the\textfont2 }
\def\{{\relax\ifmmode\lbrace\else
    {\textfontii f}\spacefactor=\@m\fi}
\def\}{\relax\ifmmode\rbrace\else
    \let\@sf=\empty\ifhmode\edef\@sf{\spacefactor=\the\spacefactor}\fi
      {\textfontii g}\@sf\relax\fi}
\def\nonhmodeerr@#1{\errmessage
     {\string#1\space allowed only within text}}
\def\linebreak{\relax\ifhmode\unskip\break\else
    \nonhmodeerr@\linebreak\fi}
\def\allowlinebreak{\relax
   \ifhmode\allowbreak\else\nonhmodeerr@\allowlinebreak\fi}
\newskip\saveskip@
\def\nolinebreak{\relax\ifhmode\saveskip@=\lastskip\unskip
  \nobreak\ifdim\saveskip@>\z@\hskip\saveskip@\fi
   \else\nonhmodeerr@\nolinebreak\fi}
\def\newline{\relax\ifhmode\null\hfil\break
    \else\nonhmodeerr@\newline\fi}
\def\nonmathaerr@#1{\errmessage
     {\string#1\space is not allowed in display math mode}}
\def\nonmathberr@#1{\errmessage{\string#1\space is allowed only in math mode}}
\def\mathbreak{\relax\ifmmode\ifinner\break\else
   \nonmathaerr@\mathbreak\fi\else\nonmathberr@\mathbreak\fi}
\def\nomathbreak{\relax\ifmmode\ifinner\nobreak\else
    \nonmathaerr@\nomathbreak\fi\else\nonmathberr@\nomathbreak\fi}
\def\allowmathbreak{\relax\ifmmode\ifinner\allowbreak\else
     \nonmathaerr@\allowmathbreak\fi\else\nonmathberr@\allowmathbreak\fi}
\def\pagebreak{\relax\ifmmode
   \ifinner\errmessage{\string\pagebreak\space
     not allowed in non-display math mode}\else\postdisplaypenalty-\@M\fi
   \else\ifvmode\penalty-\@M\else\edef\spacefactor@
       {\spacefactor=\the\spacefactor}\vadjust{\penalty-\@M}\spacefactor@
        \relax\fi\fi}
\def\nopagebreak{\relax\ifmmode
     \ifinner\errmessage{\string\nopagebreak\space
    not allowed in non-display math mode}\else\postdisplaypenalty\@M\fi
    \else\ifvmode\nobreak\else\edef\spacefactor@
        {\spacefactor=\the\spacefactor}\vadjust{\penalty\@M}\spacefactor@
         \relax\fi\fi}
\def\newpage{\relax\ifvmode\vfill\penalty-\@M\else\nonvmodeerr@\newpage\fi}
\def\nonvmodeerr@#1{\errmessage
    {\string#1\space is allowed only between paragraphs}}
\def\smallpagebreak{\relax\ifvmode\smallbreak
      \else\nonvmodeerr@\smallpagebreak\fi}
\def\medpagebreak{\relax\ifvmode\medbreak
       \else\nonvmodeerr@\medpagebreak\fi}
\def\bigpagebreak{\relax\ifvmode\bigbreak
      \else\nonvmodeerr@\bigpagebreak\fi}
\newdimen\captionwidth@
\captionwidth@=\hsize
\advance\captionwidth@ by -1.5in
\def\caption#1{}
\def\topspace#1{\gdef\thespace@{#1}\ifvmode\def\next
    {\futurelet\next\topspace@}\else\def\next{\nonvmodeerr@\topspace}\fi\next}
\def\topspace@{\ifx\next\Space@\def\next. {\futurelet\next\topspace@@}\else
     \def\next.{\futurelet\next\topspace@@}\fi\next.}
\def\topspace@@{\ifx\next\caption\let\next\topspace@@@\else
    \let\next\topspace@@@@\fi\next}
 \def\topspace@@@@{\topinsert\vbox to
       \thespace@{}\endinsert}
\def\topspace@@@\caption#1{\topinsert\vbox to
    \thespace@{}\nobreak
      \smallskip
    \setbox\z@=\hbox{\noindent\ignorespaces#1\unskip}%
   \ifdim\wd\z@>\captionwidth@
   \centerline{\vbox{\hsize=\captionwidth@\noindent\ignorespaces#1\unskip}}%
   \else\centerline{\box\z@}\fi\endinsert}
\def\midspace#1{\gdef\thespace@{#1}\ifvmode\def\next
    {\futurelet\next\midspace@}\else\def\next{\nonvmodeerr@\midspace}\fi\next}
\def\midspace@{\ifx\next\Space@\def\next. {\futurelet\next\midspace@@}\else
     \def\next.{\futurelet\next\midspace@@}\fi\next.}
\def\midspace@@{\ifx\next\caption\let\next\midspace@@@\else
    \let\next\midspace@@@@\fi\next}
 \def\midspace@@@@{\midinsert\vbox to
       \thespace@{}\endinsert}
\def\midspace@@@\caption#1{\midinsert\vbox to
    \thespace@{}\nobreak
      \smallskip
      \setbox\z@=\hbox{\noindent\ignorespaces#1\unskip}%
      \ifdim\wd\z@>\captionwidth@
    \centerline{\vbox{\hsize=\captionwidth@\noindent\ignorespaces#1\unskip}}%
    \else\centerline{\box\z@}\fi\endinsert}
\mathchardef\prime@="0230
\def\prime{{{}\prime@{}}}
\def\prim@s{\prime@\futurelet\next\pr@m@s}

\def\,{\relax\ifmmode\mskip\thinmuskip\else\thinspace\fi}
\def\!{\relax\ifmmode\mskip-\thinmuskip\else\negthinspace\fi}
\def\frac#1#2{{#1\over#2}}

\def\:{\nobreak\hskip.1111em{:}\hskip.3333em plus .0555em\relax}
\def\intic@{\mathchoice{\hskip5\p@}{\hskip4\p@}{\hskip4\p@}{\hskip4\p@}}
\def\negintic@
 {\mathchoice{\hskip-5\p@}{\hskip-4\p@}{\hskip-4\p@}{\hskip-4\p@}}
\def\intkern@{\mathchoice{\!\!\!}{\!\!}{\!\!}{\!\!}}
\def\intdots@{\mathchoice{\cdots}{{\cdotp}\mkern1.5mu
    {\cdotp}\mkern1.5mu{\cdotp}}{{\cdotp}\mkern1mu{\cdotp}\mkern1mu
      {\cdotp}}{{\cdotp}\mkern1mu{\cdotp}\mkern1mu{\cdotp}}}
\newcount\intno@
\def\iint{\intno@=\tw@\futurelet\next\ints@}
\def\iiint{\intno@=\thr@@\futurelet\next\ints@}
\def\iiiint{\intno@=4 \futurelet\next\ints@}
\def\idotsint{\intno@=\z@\futurelet\next\ints@}
\def\ints@{\findlimits@\ints@@}
\newif\iflimtoken@
\newif\iflimits@
\def\findlimits@{\limtoken@false\limits@false\ifx\next\limits
 \limtoken@true\limits@true\else\ifx\next\nolimits\limtoken@true\limits@false
    \fi\fi}
\def\multintlimits@{\intop\ifnum\intno@=\z@\intdots@
  \else\intkern@\fi
    \ifnum\intno@>\tw@\intop\intkern@\fi
     \ifnum\intno@>\thr@@\intop\intkern@\fi\intop}
\def\multint@{\int\ifnum\intno@=\z@\intdots@\else\intkern@\fi
   \ifnum\intno@>\tw@\int\intkern@\fi
    \ifnum\intno@>\thr@@\int\intkern@\fi\int}
\def\ints@@{\iflimtoken@\def\ints@@@{\iflimits@
   \negintic@\mathop{\intic@\multintlimits@}\limits\else
    \multint@\nolimits\fi\eat@}\else
     \def\ints@@@{\multint@\nolimits}\fi\ints@@@}
\def\Sb{_\bgroup\vspace@
        \baselineskip=\fontdimen10 \scriptfont\tw@
        \advance\baselineskip by \fontdimen12 \scriptfont\tw@
        \lineskip=\thr@@\fontdimen8 \scriptfont\thr@@
        \lineskiplimit=\thr@@\fontdimen8 \scriptfont\thr@@
        \Let@\vbox\bgroup\halign\bgroup \hfil$\scriptstyle
            {##}$\hfil\cr}
\def\endSb{\crcr\egroup\egroup\egroup}
\def\Sp{^\bgroup\vspace@
        \baselineskip=\fontdimen10 \scriptfont\tw@
        \advance\baselineskip by \fontdimen12 \scriptfont\tw@
        \lineskip=\thr@@\fontdimen8 \scriptfont\thr@@
        \lineskiplimit=\thr@@\fontdimen8 \scriptfont\thr@@
        \Let@\vbox\bgroup\halign\bgroup \hfil$\scriptstyle
            {##}$\hfil\cr}
\def\endSp{\crcr\egroup\egroup\egroup}
\def\Let@{\relax\iffalse{\fi\let\\=\cr\iffalse}\fi}
\def\vspace@{\def\vspace##1{\noalign{\vskip##1 }}}
\def\aligned{\,\vcenter\bgroup\vspace@\Let@\openup\jot\m@th\ialign
  \bgroup \strut\hfil$\displaystyle{##}$&$\displaystyle{{}##}$\hfil\crcr}
\def\endaligned{\crcr\egroup\egroup}
\def\matrix{\,\vcenter\bgroup\Let@\vspace@
    \normalbaselines
  \m@th\ialign\bgroup\hfil$##$\hfil&&\quad\hfil$##$\hfil\crcr
    \mathstrut\crcr\noalign{\kern-\baselineskip}}
\def\endmatrix{\crcr\mathstrut\crcr\noalign{\kern-\baselineskip}\egroup
                \egroup\,}
\newtoks\hashtoks@
\hashtoks@={#}
\def\format{\crcr\egroup\iffalse{\fi\ifnum`}=0 \fi\format@}
\def\format@#1\\{\def\preamble@{#1}%
  \def\c{\hfil$\the\hashtoks@$\hfil}%
  \def\r{\hfil$\the\hashtoks@$}%
  \def\l{$\the\hashtoks@$\hfil}%
  \setbox\z@=\hbox{\xdef\Preamble@{\preamble@}}\ifnum`{=0 \fi\iffalse}\fi
   \ialign\bgroup\span\Preamble@\crcr}

\def\cases{\left\{\,\vcenter\bgroup\vspace@
     \normalbaselines\openup\jot\m@th
       \Let@\ialign\bgroup$##$\hfil&\quad$##$\hfil\crcr
      \mathstrut\crcr\noalign{\kern-\baselineskip}}

\newif\iftagsleft@
\tagsleft@true
\def\TagsOnRight{\global\tagsleft@false}
\def\tag#1$${\iftagsleft@\leqno\else\eqno\fi
 \hbox{\def\pagebreak{\global\postdisplaypenalty-\@M}%
 \def\nopagebreak{\global\postdisplaypenalty\@M}\rm(#1\unskip)}%
  $$\postdisplaypenalty\z@\ignorespaces}
\interdisplaylinepenalty=\@M
\def\allowdisplaybreak@{\def\allowdisplaybreak{\noalign{\allowbreak}}}
\def\displaybreak@{\def\displaybreak{\noalign{\break}}}
\def\align#1\endalign{\def\tag{&}\vspace@\allowdisplaybreak@\displaybreak@
  \iftagsleft@\lalign@#1\endalign\else
   \ralign@#1\endalign\fi}
\def\ralign@#1\endalign{\displ@y\Let@\tabskip\centering\halign to\displaywidth
     {\hfil$\displaystyle{##}$\tabskip=\z@&$\displaystyle{{}##}$\hfil
       \tabskip=\centering&\llap{\hbox{(\rm##\unskip)}}\tabskip\z@\crcr
             #1\crcr}}
\def\lalign@
 #1\endalign{\displ@y\Let@\tabskip\centering\halign to \displaywidth
   {\hfil$\displaystyle{##}$\tabskip=\z@&$\displaystyle{{}##}$\hfil
   \tabskip=\centering&\kern-\displaywidth
        \rlap{\hbox{(\rm##\unskip)}}\tabskip=\displaywidth\crcr
               #1\crcr}}
\def\overrightarrow{\mathpalette\overrightarrow@}
\def\overrightarrow@#1#2{\vbox{\ialign{$##$\cr
    #1{-}\mkern-6mu\cleaders\hbox{$#1\mkern-2mu{-}\mkern-2mu$}\hfill
     \mkern-6mu{\to}\cr
     \noalign{\kern -1\p@\nointerlineskip}
     \hfil#1#2\hfil\cr}}}
\def\overleftarrow{\mathpalette\overleftarrow@}
\def\overleftarrow@#1#2{\vbox{\ialign{$##$\cr
     #1{\leftarrow}\mkern-6mu\cleaders\hbox{$#1\mkern-2mu{-}\mkern-2mu$}\hfill
      \mkern-6mu{-}\cr
     \noalign{\kern -1\p@\nointerlineskip}
     \hfil#1#2\hfil\cr}}}
\def\overleftrightarrow{\mathpalette\overleftrightarrow@}
\def\overleftrightarrow@#1#2{\vbox{\ialign{$##$\cr
     #1{\leftarrow}\mkern-6mu\cleaders\hbox{$#1\mkern-2mu{-}\mkern-2mu$}\hfill
       \mkern-6mu{\to}\cr
    \noalign{\kern -1\p@\nointerlineskip}
      \hfil#1#2\hfil\cr}}}
\def\underrightarrow{\mathpalette\underrightarrow@}
\def\underrightarrow@#1#2{\vtop{\ialign{$##$\cr
    \hfil#1#2\hfil\cr
     \noalign{\kern -1\p@\nointerlineskip}
    #1{-}\mkern-6mu\cleaders\hbox{$#1\mkern-2mu{-}\mkern-2mu$}\hfill
     \mkern-6mu{\to}\cr}}}
\def\underleftarrow{\mathpalette\underleftarrow@}
\def\underleftarrow@#1#2{\vtop{\ialign{$##$\cr
     \hfil#1#2\hfil\cr
     \noalign{\kern -1\p@\nointerlineskip}
     #1{\leftarrow}\mkern-6mu\cleaders\hbox{$#1\mkern-2mu{-}\mkern-2mu$}\hfill
      \mkern-6mu{-}\cr}}}
\def\underleftrightarrow{\mathpalette\underleftrightarrow@}
\def\underleftrightarrow@#1#2{\vtop{\ialign{$##$\cr
      \hfil#1#2\hfil\cr
    \noalign{\kern -1\p@\nointerlineskip}
     #1{\leftarrow}\mkern-6mu\cleaders\hbox{$#1\mkern-2mu{-}\mkern-2mu$}\hfill
       \mkern-6mu{\to}\cr}}}
\def\sqrt#1{\radical"270370 {#1}}
\def\dots{\relax\ifmmode\let\next=\ldots\else\let\next=\tdots@\fi\next}
\def\tdots@{\unskip\ \tdots@@}
\def\tdots@@{\futurelet\next\tdots@@@}
\def\tdots@@@{$\mathinner{\ldotp\ldotp\ldotp}\,
   \ifx\next,$\else
   \ifx\next.\,$\else
   \ifx\next;\,$\else
   \ifx\next:\,$\else
   \ifx\next?\,$\else
   \ifx\next!\,$\else
   $ \fi\fi\fi\fi\fi\fi}
\def\text{\relax\ifmmode\let\next=\text@\else\let\next=\text@@\fi\next}
\def\text@@#1{\hbox{#1}}
\def\text@#1{\mathchoice
 {\hbox{\everymath{\displaystyle}\def\textfonti{\the\textfont1 }%
    \def\textfontii{\the\textfont2 }\textdef@@ T#1}}
 {\hbox{\everymath{\textstyle}\def\textfonti{\the\textfont1 }%
    \def\textfontii{\the\textfont2 }\textdef@@ T#1}}
 {\hbox{\everymath{\scriptstyle}\def\textfonti{\the\scriptfont1 }%
   \def\textfontii{\the\scriptfont2 }\textdef@@ S\rm#1}}
 {\hbox{\everymath{\scriptscriptstyle}\def\textfonti{\the\scriptscriptfont1 }%
   \def\textfontii{\the\scriptscriptfont2 }\textdef@@ s\rm#1}}}
\def\textdef@@#1{\textdef@#1\rm \textdef@#1\bf
   \textdef@#1\sl \textdef@#1\it}

\def\textdef@#1#2{\def\next{\csname\expandafter\eat@\string#2fam\endcsname}%
\if S#1\edef#2{\the\scriptfont\next\relax}%
 \else\if s#1\edef#2{\the\scriptscriptfont\next\relax}%
 \else\edef#2{\the\textfont\next\relax}\fi\fi}
\scriptfont\itfam=\tenit \scriptscriptfont\itfam=\tenit
\scriptfont\slfam=\tensl \scriptscriptfont\slfam=\tensl
\mathcode`\0="0030
\mathcode`\1="0031
\mathcode`\2="0032
\mathcode`\3="0033
\mathcode`\4="0034
\mathcode`\5="0035
\mathcode`\6="0036
\mathcode`\7="0037
\mathcode`\8="0038
\mathcode`\9="0039
\def\Cal{\relax\ifmmode\let\next=\Cal@\else
     \def\next{\errmessage{Use \string\Cal\space only in math mode}}\fi\next}
\def\Cal@#1{{\fam2 #1}}
\def\bold{\relax\ifmmode\let\next=\bold@\else
   \def\next{\errmessage{Use \string\bold\space only in math
      mode}}\fi\next}\def\bold@#1{{\fam\bffam #1}}
\mathchardef\Gamma="0000
\mathchardef\Delta="0001
\mathchardef\Theta="0002
\mathchardef\Lambda="0003
\mathchardef\Xi="0004
\mathchardef\Pi="0005
\mathchardef\Sigma="0006
\mathchardef\Upsilon="0007
\mathchardef\Phi="0008
\mathchardef\Psi="0009
\mathchardef\Omega="000A
\mathchardef\varGamma="0100
\mathchardef\varDelta="0101
\mathchardef\varTheta="0102
\mathchardef\varLambda="0103
\mathchardef\varXi="0104
\mathchardef\varPi="0105
\mathchardef\varSigma="0106
\mathchardef\varUpsilon="0107
\mathchardef\varPhi="0108
\mathchardef\varPsi="0109
\mathchardef\varOmega="010A
\font\dummyft@=dummy
\fontdimen1 \dummyft@=\z@
\fontdimen2 \dummyft@=\z@
\fontdimen3 \dummyft@=\z@
\fontdimen4 \dummyft@=\z@
\fontdimen5 \dummyft@=\z@
\fontdimen6 \dummyft@=\z@
\fontdimen7 \dummyft@=\z@
\fontdimen8 \dummyft@=\z@
\fontdimen9 \dummyft@=\z@
\fontdimen10 \dummyft@=\z@
\fontdimen11 \dummyft@=\z@
\fontdimen12 \dummyft@=\z@
\fontdimen13 \dummyft@=\z@
\fontdimen14 \dummyft@=\z@
\fontdimen15 \dummyft@=\z@
\fontdimen16 \dummyft@=\z@
\fontdimen17 \dummyft@=\z@
\fontdimen18 \dummyft@=\z@
\fontdimen19 \dummyft@=\z@
\fontdimen20 \dummyft@=\z@
\fontdimen21 \dummyft@=\z@
\fontdimen22 \dummyft@=\z@
\def\fontlist@{\\{\tenrm}\\{\sevenrm}\\{\fiverm}\\{\teni}\\{\seveni}%
 \\{\fivei}\\{\tensy}\\{\sevensy}\\{\fivesy}\\{\tenex}\\{\tenbf}\\{\sevenbf}%
 \\{\fivebf}\\{\tensl}\\{\tenit}\\{\tensmc}}
\def\dodummy@{{\def\\##1{\global\let##1=\dummyft@}\fontlist@}}
\newif\ifsyntax@
\newcount\countxviii@
\def\newtoks@{\alloc@5\toks\toksdef\@cclvi}
\def\nopages@{\output={\setbox\z@=\box\@cclv \deadcycles=\z@}\newtoks@\output}
\def\syntax{\syntax@true\dodummy@\countxviii@=\count18
\loop \ifnum\countxviii@ > \z@ \textfont\countxviii@=\dummyft@
   \scriptfont\countxviii@=\dummyft@ \scriptscriptfont\countxviii@=\dummyft@
     \advance\countxviii@ by-\@ne\repeat
\dummyft@\tracinglostchars=\z@
  \nopages@\frenchspacing\hbadness=\@M}
\def\magstep#1{\ifcase#1 1000\or
 1200\or 1440\or 1728\or 2074\or 2488\or
 \errmessage{\string\magstep\space only works up to 5}\fi\relax}
{\lccode`\2=`\p \lccode`\3=`\t
 \lowercase{\gdef\tru@#123{#1truept}}}

\def\scaletype#1{\mag=#1\relax
 \hsize=\expandafter\tru@\the\hsize
 \vsize=\expandafter\tru@\the\vsize
 \dimen\footins=\expandafter\tru@\the\dimen\footins}

\def\scalefont#1#2\andcallit#3{\edef\font@{\the\font}#1\font#3=
  \fontname\font\space scaled #2\relax\font@}
\def\Mag@#1#2{\ifdim#1<1pt\multiply#1 #2\relax\divide#1 1000 \else
  \ifdim#1<10pt\divide#1 10 \multiply#1 #2\relax\divide#1 100\else
  \divide#1 100 \multiply#1 #2\relax\divide#1 10 \fi\fi}
\def\scalelinespacing#1{\Mag@\baselineskip{#1}\Mag@\lineskip{#1}%
  \Mag@\lineskiplimit{#1}}
\def\wlog#1{\immediate\write-1{#1}}
\catcode`\@=\active
\mag=\magstep1
\hsize=162truemm
\vsize=230truemm
\baselineskip=8truemm
\nopagenumbers
\pageno=1
\footline={\hss - {\it\folio} - \hss}
\TagsOnRight
\overfullrule=0pt
\def\=def{\; \mathop{=}_{\text{\rm def}} \;}
\def\rd{\partial}

\def\Z{{\bold Z}}
\def\calB{{\cal B}}
\def\calL{{\cal L}}
\def\calM{{\cal M}}
\def\calU{{\cal U}}
\def\calV{{\cal V}}
\def\calP{{\cal P}}
\def\calQ{{\cal Q}}

\def\calUhat{{\hat{\calU}}}
\def\calVhat{{\hat{\calV}}}

\def\pbar{{\bar{p}}}
\def\qbar{{\bar{q}}}
\def\zbar{{\bar{z}}}
\def\SDiff(2){ \text{SDiff(2)} }
\def\zetahat{ \hat{\zeta} }
%
%
%
\line{{\it College of Liberal Arts and Sciences}
  \hfill KUCP-0049/92}
\line{{\it Kyoto University}
  \hfill revised version June 1992}

\title
    W ALGEBRA, TWISTOR, AND \\
    NONLINEAR INTEGRABLE SYSTEMS%
    \footnote"$^\sharp$"
      {\baselineskip=10pt
       Expanded version of talk at RIMS workshop
       ``Algebraic Analysis,'' Kyoto University,
       March 23-26, 1992}
    \\
\endtitle
\author
    Kanehisa Takasaki\footnote"$^\flat$"%
      {
       E-mail: takasaki\@jpnyitp.bitnet, %
               takasaki\@kurims.kyoto-u.ac.jp%
      }\\
    Institute of Mathematics, Yoshida College, Kyoto University\\
    {\it Yoshida-Nihonmatsu-cho, Sakyo-ku, Kyoto 606, Japan}\\
\endauthor

\vskip15truemm

\noindent
{\bf Abstract}. W algebras arise in the study of various nonlinear
integrable systems such as: self-dual gravity, the KP and Toda
hierarchies, their quasi-classical (or dispersionless) limit, etc.
Twistor theory provides a geometric background for these algebras.
Present state of these topics is overviewed. A few ideas on possible
deformations of self-dual gravity (including quantum deformations)
are presented.
\vfill
\eject
\heading    1. Introduction
\endheading

\noindent
The dramatic progress of the 2D gravity/string theory
in the last few years [1] has revealed a new relation of
field theory to integrable hierarchies of KdV, KP and
Toda lattice type.
The theory of nonlinear integrable systems has thus
again proven its usefulness in physics. It is Virasoro
and W symmetries of these integrable hierarchies that
plays a central role in characterizing these models of
gravity and strings as special solutions of an integrable
hierarchy.

The structure of Virasoro and W symmetries arises in
several different (but actually equivalent) forms.
Firstly, the partition function $Z$ (or its square root
$\sqrt{Z}$) of these models can be identified with the
tau function $\tau$ of an integrable hierarchy, and
satisfies a set of linear constraints (Virasoro constraints)
of the form
$$
    \calL_n \tau = 0, \quad n \ge -1                \tag 1
$$
for a set of Virasoro symmetry generators $\calL_n$; these
constraints can be further generalized to W algebraic analogues
(W constraints)[2]. Secondly, the same model of 2D gravity
can be reproduced from the canonical commutation relation
(Douglas equation) [3]
$$
    [P, Q] = 1                                      \tag 2
$$
of two ordinary differential operators in one variable.
A third expression is due to the Schwinger-Dyson equation (or loop
equations) for loop correlation functions [4]. Although less obvious,
the latter two expressions of 2D gravity, too, stem from Virasoro
and W symmetries.

W symmetries also exist in the self-dual vacuum Einstein equation [5],
a higher dimensional nonlinear integrable system.  Not only being
an integrable model of 4D gravity (self-dual gravity), this equation
(as well as its hyper-K\"ahler versions in 4k dimensions) has also
been extensively studied in the context of supersymmetric nonlinear
sigma models [6], relativistic membranes [7], SU($\infty)$ Toda
fields [8] etc., and very recently, as an effective theory of $N=2$
strings [9]. These diverse models of field theory may be thought of
as higher dimensional counterparts of the above mentioned models of
the 2D gravity/string theory.

Our basic standpoint is that W symmetries (in particular,
$W_{1+\infty}$, $w_{1+\infty}$ and their variations [10]) provide
us with a unified framework for understanding these nonlinear
integrable systems. We start with a brief review of W algebraic
structures in self-dual gravity, then turn to similar results on
the KP and Toda hierarchies and their quasi-classical limits.
In the final section, we shall present a few ideas on deformations
of self-dual gravity and associated W algebraic structures.

\heading    2. Self-dual gravity
\endheading

\noindent
W algebraic structures of self-dual gravity can be deduced from
Penrose's twistor theoretical approach (nonlinear graviton
construction) [11].  To see this, it is convenient to start from
the Plebanski equation [12]
$$
    \Omega_{,p\pbar}\Omega_{,q\qbar}
    - \Omega_{,p\qbar}\Omega_{,q\pbar} = 1          \tag 3
$$
where $\Omega$ is a K\"ahler potential and $p,q,\pbar,\qbar$
are suitably chosen complex coordinates.  This equation
(actually known to mathematicians before Plebanski) represents
Ricci-flatness of a K\"ahler metric. We now introduce a new
variable $\lambda$ (known in the theory of nonlinear integrable
systems as ``spectral parameter'') and, following Gindikin [13],
make a linear combination
$$
    \omega(\lambda) \=def d\pbar \wedge d\qbar +\lambda \omega
                      + \lambda^2 dp \wedge dq   \tag 4
$$
of the holomorphic 2-form $dp \wedge dq$, the anti-holomorphic
2-form $d\pbar \wedge d\qbar$ and the K\"ahler form
$\omega \=def \Omega_{,p^i \pbar^j} dp^i \wedge d\pbar^j$,
$p^i =(p,q)$, $\pbar^i = (\pbar,\qbar)$.  The Plebanski equation
can be now cast into the exterior differential equations
$$
    d\omega(\lambda) = 0, \quad
    \omega(\lambda) \wedge \omega(\lambda) = 0 \quad
    (d\lambda = 0),                                     \tag 5
$$
where $d$ stands for total differential in $(p,q,\pbar,\qbar)$
viewing $\lambda$ a constant.

By a classical theorem of Darboux, one can find two ``Darboux
coordinates'' $P(\lambda)$ and $Q(\lambda)$ as
$$
    \omega(\lambda) = d P(\lambda) \wedge d Q(\lambda) \quad
    (d\lambda = 0).                                     \tag 6
$$
In particular, these Darboux coordinates give a canonical
conjugate pair
$$
    \{ P(\lambda), Q(\lambda) \}_{\pbar,\qbar} = 1      \tag 7
$$
for the Poisson bracket $ \{ F, G \}_{\pbar,\qbar}
\=def F_{,\pbar}G_{,\qbar} - F_{,\qbar}G_{,\pbar}$.
Actually, these Darboux coordinates are not unique,
but allow transformations
$$
    P(\lambda),Q(\lambda) \ \longrightarrow \
    f\bigl( \lambda,P(\lambda),Q(\lambda) \bigr),
    g\bigl( \lambda,P(\lambda),Q(\lambda) \bigr)        \tag 8
$$
by a two-dimensional symplectic (i.e, area-preserving) diffeomorphism
depending also on $\lambda$; $f$ and $g$ are required to have a unit
Jacobian for the second and third variables, hence defines an
area-preserving diffeomorphism with parameter $\lambda$.
The relevance of a W algebra (in this case, $w_{1+\infty}$)
is already manifest.

Penrose's idea is to consider two special pairs of Darboux
coordinates, say $\calU(\lambda),\calV(\lambda)$ and
$\calUhat(\lambda),\calVhat(\lambda)$, with different complex
analytic properties with respect to $\lambda$.  These Darboux
coordinate systems are then linked with each other by a symplectic
mixing as in (8):
$$
    f\bigl( \lambda,\calU(\lambda),\calV(\lambda) \bigr)
      = \calUhat(\lambda),                               \quad
    g\bigl( \lambda,\calU(\lambda),\calV(\lambda) \bigr)
      = \calVhat(\lambda),                               \tag 9
$$
and this gives a Riemann-Hilbert problem in the group SDiff(2) of
area preserving diffeomorphisms.  The data $(f,g)$, which then
becomes an element of the loop group $\calL\SDiff(2)$ of SDiff(2),
is exactly Penrose's twistor data, and conversely, solving the above
Riemann-Hilbert problem (which is generally a hard task though)
for a given data give rise to all (local) solutions of self-dual
gravity .

The existence of a large set of symmetries is now an obvious
consequence of the $\calL\SDiff(2)$ group structure in the twistor
data $(f,g)$: the action of this loop group on itself (from
left or right) gives transformations of the corresponding solution
of self-dual gravity via the Riemann-Hilbert problem.
Infinitesimal symmetries accordingly have the structure of the loop
algebra of $w_{1+\infty}$.

Algebraic structures found in self-dual gravity are thus
more or less reminiscent of 2D gravity as well as W gravity [14],
in which W algebras (both quantum and quasi-classical) give basic
symmetries.  This is also the case for dimensionally reduced models
of 4D self-dual gravity [15]. Note, however, that the full symmetry
algebra of 4D self-dual gravity is the loop algebra of $w_{1+\infty}$,
far larger than $w_{1+\infty}$ itself. This reflects a higher
dimensional characteristic of self-dual gravity, and suggests a
possible direction of extending the notion of W algebras. We shall
return to this issue in the final section.

\heading    3. KP hierarchy and canonical conjugate pair
\endheading

\noindent
We have seen that a canonical conjugate pair, $\calU$ and $\calV$,
takes place in the description of general solutions of self-dual
gravity. The KP hierarchy, too, turns out to have a similar pair
(of pseudo-differential operators).

The KP hierarchy, by definition, describes a commuting set of
isospectral flows
$$
    \frac{\rd L}{\rd t_n} = [ B_n, L], \quad
    B_n \=def (L^n)_{\ge 0}, \quad
    n=1,2,\ldots,                                \tag 10
$$
of a one-dimensional pseudo-differential operator
$$
    L \=def \rd + \sum_{n=1}^\infty u_{n+1}\rd^{-n}, \quad
    \rd \=def \rd/\rd x,                         \tag 11
$$
where $(\quad)_{\ge 0}$ stands for dropping negative powers
of $\rd$ to obtain a differential operator.  This is the
ordinary Lax formalism of the KP hierarchy.

We need some other variables to describe the $W_{1+\infty}$
symmetries explicitly.  One way is to use the tau function to
realize those symmetries as linear differential operators
$W^{(s)}_n$, $s=1,2,\ldots$, $n \in \Z$,  in the $t$'s [16].
Another way is to introduce a pseudo-differential
operator of the form
$$
    M = \sum_{n=1}^\infty nt_n L^{n-1} + O(\rd^{-1}) \tag 12
$$
that satisfy the Lax equations
$$
    \frac{\rd M}{\rd t_n} = [ B_n, M ], \quad
    n=1,2,\ldots,                                  \tag 13
$$
and the canonical commutation relation
$$
    [ L, M ] = 1.                                  \tag 14
$$
Such a second Lax operator does exists, and arises in the linear
system of the so called Baker-Akhiezer function
$\psi = \psi(x,t,\lambda)$ as:
$$
    L\psi = \lambda \psi, \quad
    M\psi = \frac{\rd \psi}{\rd \lambda}, \quad
    \frac{\rd \psi}{\rd t_n} = B_n \psi.           \tag 15
$$
The $W_{1+\infty}$ symmetries of the KP hierarchy can be
reformulated as symmetries acting on this $(L,M)$ pair [17].

The above description of $W_{1+\infty}$ symmetries elucidates an
origin of $(P,Q)$ pairs in $d\le 1$ string theory [18]. The
Douglas pair for $d<1$ strings, indeed, is given by a
(noncommutative) canonical transformation
$$
    P = L^p, \quad Q = ML^{1-p}/p + h(L)           \tag 16
$$
of the $(L,M)$ pair under the constraints
$$
    P = (P)_{\ge 0}, \quad Q = (Q)_{\ge 0},         \tag 17
$$
where $h(L) = \sum h_nL^n$ with suitable constant coefficients $h_n$.
At the $(p,q)$ critical point, the time variables are restricted
to so called ``small phase space": $t_{p+q}=p/(p+q)$,
$t_{p+q+1}=t_{p+q+2}= \cdots = 0$.

\heading    4. Quasi-classical limit of KP hierarchy
\endheading

\noindent
The KP hierarchy has a quasi-classical (or dispersionless) limit [19].
This is a system of Lax type,
$$
    \frac{\rd \calL}{\rd t_n} = \{ \calB_n, \calL \}_{k,x}, \quad
    \calB_n = (\calL^n)_{\ge 0}, \quad
    n=1,2,\ldots,                                   \tag 18
$$
where $\calL$, a quasi-classical counterpart of $L$, is a Laurent
series of the form
$$
    \calL \=def k + \sum_{n=1}^\infty u_{n+1} k^{-n},   \tag 19
$$
$k$ is a parameter like $\lambda$, $(\quad)_{\ge 0}$ now
means dropping all negative powers of $k$, and $\{\quad,\quad\}_{k,x}$
the Poisson bracket in $(k,x)$:
$\{ F, G \}_{k,x} \=def F_{,k}G_{,x} - F_{,x}G_{,k}$.
As in the case of the KP hierarchy, one can introduce a second
Laurent series $\calM = \sum_{n=1}^\infty nt_n\calL^{n-1} + O(k^{-1})$
that obeys similar equations,
$$
    \frac{\rd \calM}{\rd t_n} = \{ \calB_n, \calM \}_{k,x}, \quad
    \{ \calL, \calM \}_{k,x} = 1.                        \tag 20
$$

The above hierarchy (dispersionless or quasi-classical KP hierarchy)
gives a quasi-classical limit of the KP hierarchy in the following
sense.  Introduce a Planck constant $\hbar$ into the KP hierarchy
and the associated linear system by replacing
$$
    \rd=\frac{\rd}{\rd x} \to \hbar\frac{\rd}{\rd x}, \quad
    \frac{\rd}{\rd t_n} \to \hbar\frac{\rd}{\rd t_n}, \quad
    \frac{\rd}{\rd \lambda} \to \hbar\frac{\rd}{\rd \lambda}
                                                        \tag 21
$$
and assume a quasi-classical (WKB) asymptotic form of the
the Baker-Akhiezer function,
$$
    \psi(\hbar,x,t,\lambda)
      \sim \exp \hbar^{-1} S(x,t,\lambda).              \tag 22
$$
The linear system then gives rise to a set of eikonal
(or Hamilton-Jacobi) equations for the phase function $S$,
which after somewhat lengthy calculations turn out to be
equivalent to the above equations of the $(\calL,\calM)$ pair.
Futher, the tau function has accordingly an asymptotic form
$$
    \tau \sim \exp[ \hbar^{-2}F(x,t) + O(\hbar^{-1}) ].      \tag 23
$$
In view of the relation to matrix models of 2D gravity in large-$N$
limit [1], the function $F$ should be called the ``free energy" of the
quasi-classical KP hierarchy. Its exponential $\exp F$ gives
exactly the tau function introduced in Ref. 20. A set of
$w_{1+\infty}$ ($=$ SDiff(2)) symmetries are also constructed
in the same paper.

A quasi-classical limit of the Douglas pair $(P,Q)$ is given by
$$
    \calP \=def \calL^p, \quad
    \calQ \=def \calM\calL^{1-p}/p + h(\calL),       \tag 24
$$
and constrained by
$$
    \calP = (\calP)_{\ge 0}, \quad
    \calQ = (\calQ)_{\ge 0}.                         \tag 25
$$
The ``genus zero'' part of 2D gravity [1] as well as ``topological
minimal models'' [21] are included into this family of solutions.

\heading    5. Toda lattice and its quasi-classical limit
\endheading

\noindent
What we have seen in the previous two sections persists in
the Toda lattice hierarchy. The relativistic Toda field theory
is given by the equation of motion
$$
    \frac{\rd^2 \Phi_n}{\rd z \rd \zbar} + \exp(\Phi_{n+1}-\Phi_n)
     - \exp(\Phi_n - \Phi_{n-1}) = 0, \quad n \in \Z.   \tag 26
$$
In quasi-classical limit, the discrete variable $n$, too, has to
be scaled as $\hbar n = s$ [22], and one obtains the equation
$$
    \frac{\rd^2 \Phi}{\rd z \rd \zbar}
    + \frac{\rd}{\rd s}\exp\frac{\rd\Phi}{\rd s} = 0    \tag 27
$$
for a three-dimensional field $\Phi=\Phi(z,\zbar,s)$. Because of
this correspondence, the above equation [also called SU($\infty$)
Toda equation] has been studied in detail by the methods of conformal
field theories [8] and nonlinear integrable systems [23].
The notion of tau function (or free energy) and $w_{1+\infty}$
symmetries, too, have been established [24].

Quite accidentally, the same equation describes a dimensional
reduction of self-dual gravity by a rotational $S^1$ symmetry,
as first pointed out by relativists [25].  In their interpretation,
remarkably, the $\Phi$ field is nothing but the radial coordinate
of a 4D cylindrical coordinate system; a Legendre-like
transformation converts it into a dependent variable.  This is
somewhat reminiscent of the fact [26] that the Liouville mode in
a subcritical string theory can be interpreted as a time-like
coordinate in a critical string theory. The $\Phi$ field might be
a kind of Liouville mode in higher dimensional strings or membranes
(in suitable quantization, if necessary).

We have seen that the quasi-classical version of the KP/Toda
hierarchy has two distinct characteristics in itself: In one hand,
it has a Lax formalism very similar to the ordinary KP/Toda
hierarchy; on the other hand, it has a pair of canonical conjugate
variables like those in self-dual gravity.  The corresponding
twistor theory is a kind of ``minitwistor theory" [27] associated
with a two (rather than three) complex dimensional twistor space.
Naturally, one may imagine that a higher dimensional analogue of
the KP/Toda hierarchy should exist and reproduce self-dual gravity
as a kind of quasi-classical limit.  We now turn to this issue.

\heading    6. Deformations of self-dual gravity
\endheading

\noindent
The ordinary twistor theoretical framework based on 3D twistor
spaces already provides us  with a wide range of deformations
of self-dual gravity.  Penrose's nonlinear graviton construction,
indeed, covers all conformally self-dual spaces.  An interesting
subfamily of deformations describing an Einstein-Maxwell theory
is proposed by Flaherty [28] and recently studied by Park [29].
It recently turned out that a group of volume-preserving
diffeomorphisms, SDiff(3), underlies this family of deformations
and plays the same role as the $\calL\SDiff(2)$ group in self-dual
gravity [30]. This kind of deformations associated with an SDiff(3)
group deserve further study.

Another idea, which might lead to quantum deformations, is to
generalize the correspondence between the KP hierarchy and its
quasi-classical version to self-dual gravity. As already mentioned,
the correspondence
$$
    \text{KP/Toda hierarchy} \
    \mathop{\longrightarrow}_{\hbar \to 0} \
    \text{quasi-classical KP/Toda hierarchy}
$$
strongly suggests a higher dimensional analogue such as
$$
    \text{\quad ? \quad}
    \mathop{\longrightarrow}_{\hbar \to 0} \
    \text{4D self-dual gravity}
$$
At the place of ``?'' should come a kind of quantization of
self-dual gravity and twistor theory.
A symmetry algebra coming into the place of ``?'' should be a
quantum deformation of the loop algebra $\calL w_{1+\infty}$
of $w_{1+\infty}$.

A candidate of quantum deformations of $\calL w_{1+\infty}$
is the loop algebra $\calL W_{1+\infty}$ of $W_{1+\infty}$.
We do not know what an associated deformation of self-dual
gravity looks like.  Recent diverse proposals for a 3D field
formulation of $d=1$ matrix models [31] are very suggestive
in that respect.  Also interesting are a family of nonlinear
integrable systems recently presented by Hoppe et al [32];
Lax representations of these models exploit the $W_{1+\infty}$
algebra or the Moyal algebra [33] in a quite explicit way. This
will also be related to the star-product membrane theory [34],
a deformation of relativistic membrane theory with a Moyal
bracket replacing a Poisson bracket in its Hamiltonian density.
Such a possible link with membrane theory is very significant,
because Ooguri and Vafa [9] point out that $N=2$ strings look
like membranes.

Another possible deformation of $\calL w_{1+\infty}$  might be due
to the notion of ``quantized spectral parameters" [35]. A basic
idea of this notion is, roughly, to replace a spectral (i.e., loop)
parameter, say $\zeta$, by an operator of the form
$$
    \zetahat = \zeta \exp(-\hbar \rd/\rd x),
$$
where $x$ is a space variable like that of the KP hierarchy.
One can indeed derive such an operator from a reduction of
the Toda lattice (or, rather, modified KP) hierarchy [36].
If the loop algebra $\calL w_{1+\infty}$ can be deformed to a
``quantized'' loop algebra with such a ``quantized spectral
parameter," an associated deformation of self-dual gravity,
if exists, will naturally include the extra variable $x$
within its independent variables.  This is a very interesting
possibility, because the deformed self-dual gravity then will
have a direct connection with a KP-type hierarchy, hence will
offer a hint to unify self-dual gravity with KP-type hierarchies.

A similar idea of deformations of self-dual gravity can be found
in a paper of Bakas and Kiritsis [37].  They first introduce an
extension $W^N_\infty$ of $W_\infty$ (or rather $W_{1+\infty}$)
with U($N$) inner symmetries, and point out that the large-$N$
limit $W^\infty_\infty$ of $W^N_\infty$ will become isomorphic to
the SpDiff(4) algebra of 4D infinitesimal symplectic diffeomorphisms.
(In the current convention of W algebra, therefore, this algebra
should rather be called $w^\infty_\infty$; no quantum deformation
seems to have been constructed until now.)  This algebra includes
the loop algebra $\calL w_{1+\infty}$ of $w_{1+\infty}$ (i.e., of
the SDiff(2) algebra) as a subalgebra. On the basis of these
observations, they argue that this algebra (and possible quantum
deformations) should be related to a quantum deformation of self-dual
gravity.

Note that the loop algebra $\calL w_{1+\infty}$ is three dimensional
in its nature, the three variables being, e.g., $\lambda$, $\pbar$
and $\qbar$. (There are some other choices of those variables [5].)
The $W^\infty_\infty$ algebra should be accompanied with four variables,
i.e., canonical coordinates of a 4D symplectic manifold.  It is
amusing to imagine that the extra variable $x$ associated with the
quantized spectral parameter is exactly the fourth one that should
be added to the previous three variables.

This will suggest to consider an extended 4D twistor space rather
than an ordinary 3D twistor space. The first two variables $\lambda$
(or rather $k$) and $x$ are fundamental ingredients of the
quasi-classical KP hierarchy. A 4D symplectic manifold with four
coordinates, say  $\lambda,x,y,z$ and a symplectic form
$d\lambda \wedge dx + dy \wedge dz$ apparently look like a nice
framework for the aforementioned unification program.  Unfortunately,
this program has not been successful due to unexpected difficulties.
\footnote"$^*$"{An earlier version of this paper included
errors on this issue}
This is a quite technical issue and details are omitted here.

A more hopeful direction would be that of the ordinary formulation of
nonchiral $w_\infty$ gravity [14] and its hypothetical ``topological''
version [38]. In these theories, SpDiff(4) symmetries are rather
living in a 4D space-time with a symplectic structure (typically, the
cotangent bundle $T^*\Sigma$ of a Riemann surface $\Sigma$), or acting
on a moduli space $M_\infty$ of such manifolds.  The $\calL w_{1+\infty}$
subalgebra of SpDiff(4), in that picture, cannot be identified with
the twistor theoretical symmetry algebra that stems from a 3D twistor
space. Nevertheless, a link with self-dual gravity still persists as
Hitchin conjectures [38].  This conjecture seems to have been verified
in the context of the $N=2$ string theory by Ooguri and Vafa [9];
they show a construction of a hyper-K\"ahler metric (i.e.,  a solution
of self-dual gravity) on $T^*\Sigma$.  It would be interesting to see
how the SpDiff(4) algebra on $M_\infty$ act on these solutions;
those SpDiff(4) symmetries might give rise to ``constraints'' like
the W constraints of 2D gravity.  If this is true, we expect a new
nonlinear integrable system to lie behind.

\heading    References
\endheading

\item{[1]} 
For an overview, see:
Kaku, M., Strings, conformal fields, and topology
(Springer-Verlag, 1991), Chapters 13 and 14,
and references cited therein.

\item{[2]} 
Fukuma, M., Kawai, H., and Nakayama, R.,
Commun. Math. Phys. 143 (1991), 371-403.

\item{[3]} 
Douglas, M.
Phys. Lett. 238B (1990) 176-180.

\item{[4]} 
David, F.,
Mod. Phys. Lett. A 5 (1990), 1019-1029.

\item{}
Dijkgraaf, R., Verlinde, E., and Verlinde, H.,
Nucl. Phys. B348 (1991), 435-456.

\item{}
Fukuma, M., Kawai, H., and Nakayama, R.,
Int. J. Mod. Phys. A6 (1991), 1385-1406.

\item{[5]} 
Boyer, C.P., and Plebanski, J.F.,
J. Math. Phys. 26 (1985), 229-234.

\item{}
Takasaki, K.,
J. Math. Phys. 31 (1990), 1877-1888.

\item{}
Park, Q-Han,
Phys. Lett. 238B (1990), 287-290.

\item{[6]} 
Zumino, B.,
Phys. Lett. 87B (1979), 206-209.

\item{}
Alvarez-Gaum\'e, L., Freedman, D.Z.,
Phys. Lett. 94B (1980), 171.

\item{}
Hitchin, N.J., Kahlhede, A., Lindstr\"om, U., and Ro\v{c}ek, M.,
Commun. Math. Phys. 108 (1987), 535-589.

\item{[7]} 
Floratos, E.G, and Leontaris, G.K.,
Phys. Lett. 223B (1989), 153-156.

\item{[8]} 
Bakas, I.,
Commun. Math. Phys. 134 (1990), 487-508.

\item{}
Park, Q-Han,
Phys. Lett. 236B (1990), 429-432.

\item{[9]} 
Ooguri, H., and Vafa, C.,
Mod. Phys. Lett. A5 (1990), 1389-1398;
Nucl. Phys. B361 (1991), 469-518;
Nucl. Phys. B367 (1991), 83-104.

\item{[10]} 
For a review of W algebras and symmetries in field theory, see:
Sezgin, E.,
Aspects of $W_\infty$ symmetry,
Texas A\& M preprint CTP-TAMU-9/91;
Area-preserving diffeomorphisms, $w_\infty$
algebras and $w_\infty$ gravity,
Texas A\& M preprint CTP-TAMU-13/92.

\item{[11]} 
Penrose, R.,
Gen. Rel. Grav. 7 (1976), 31-52.

\item{[12]} 
Plebanski, J.F.,
J. Math. Phys. 16 (1975), 2395-2402.

\item{[13]} 
Gindikin, S.G.,
in:
H.D. Doebner and T. Weber (eds.),
{\it Twistor Geometry and Non-linear Systems},
Lecture Notes in Mathematics  vol. 970
(Springer-Verlag, 1982).

\item{[14]} 
Bergshoeff, E., Pope, C.N., Romans, L.J., Sezgin, E., Shen, X.,
and Stelle, K.S.,
Phys. Lett. 243B (1990), 350-357.

\item{}
Bergshoeff, E., and Pope, C.N.,
Phys. Lett. 249B (1990), 208-215.

\item{[15]} 
Park, Q-Han,
in Ref. 8.

\item{}
Yamagishi, K., and Chapline, F.,
Class. Quantum Grav. 8 (1991), 427-446.

\item{}
Hull, C.M.,
Phys. Lett. 269B (1991), 257-263.

\item{[16]} 
Date, E., Kashiwara, M., Jimbo, M., and Miwa, T.,
in:
M. Jimbo and T. Miwa (eds.),
 {\it Nonlinear Integrable Systems --
Classical Theory and Quantum Theory}
(World Scientific, Singapore, 1983).

\item{}
Sato, M., and Sato, Y.,
in:
P.D. Lax, H. Fujita, and G. Strang (eds.),
{\it Nonlinear Partial Differential Equations in Applied Sciences}
(North-Holland, Amsterdam, and Kinokuniya, Tokyo, 1982).

\item{[17]} 
Orlov, A.Yu.,
in:
V.G. Bar'yakhtar et al. (eds.),
{\it Plasma Theory and Nonlinear and Turbulent Processes in Physics}
(World Scientific, Singapore, 1987).

\item{}
Grinevich, P.G., and Orlov, A.Yu.,
in:
A. Belavin et al. (eds.),
 {\it Problems of modern quantum field theory\/}
(Springer-Verlag, 1989).

\item{[18]} 
Awada, M., and Sin, S.J.,
Twisted $W_\infty$ symmetry of the KP hierarchy
and the string equation of $d=1$ matrix models,
Florida preprint UFITFT-HEP-90-33.

\item{}
Yoneya, T.,
Toward a canonical formalism of non-perturbative
two-dimensional gravity,
UT-Komaba 91-8 (February, 1991).

\item{[19]} 
Kodama, Y.,
Phys. Lett. 129A (1988), 223-226.

\item{}
Kodama, Y., and Gibbons, J.,
Phys. Lett. 135A (1989), 167-170.

\item{}
Krichever, I.M.,
Commun. Math. Phys. 143 (1991), 415-426.

\item{[20]} 
Takasaki, K., and Takebe, T.,
SDiff(2) KP hierarchy,
Kyoto preprint RIMS-814
(October, 1991; revised version December, 1991).

\item{[21]} 
Dijkgraaf, R., Verlinde, H., and Verlinde, E.,
Topological strings in $d<1$,
PUPT-1204, IASSNS-HEP-9o/71 (October, 1990).

\item{}
Blok, B., and Varchenko, A.,
Topological conformal field theories
and flat coordinates,
IASSNS-HEP-91/05 (January, 1990).

\item{[22]} 
Kodama, Y.,
Phys. Lett. 147A (1990), 477-482.

\item{[23]} 
Golenisheva-Kutuzova, M.I., and Reiman, A.G.,
Zap. Nauch. Semin. LOMI 169 (1988), 44. (in Russian).

\item{}
Saveliev, M.V., and Vershik, A.M.,
Commun. Math. Phys. 126 (1989), 367-378.

\item{}
Kashaev, R.M., Saveliev, M.V., Savelieva, S.A., and Vershik, A.M.,
in: {\it Ideas and Methods in Mathematics and Physics}
(Cambridge University Press, Cambridge, 1991).

\item{[24]} 
Takasaki, K., and Takebe, T.,
Lett. Math. Phys. 23 (1991), 205-214.

\item{[25]} 
Boyer, C., and Finley, J.D.,
J. Math. Phys. 23 (1982), 1126-1128.

\item{}
Gegenberg, J.D., and Das, A.,
Gen. Rel. Grav. 16 (1984), 817-829.

\item{[26]} 
Dhar, A., Jayaraman, T., Narain, K.S., and Wadia, S.R.,
Mod. Phys. Lett. A5 (1990), 863.

\item{[27]} 
Hitchin, N.J.,
in:
H.D. Doebner and T. Weber (eds.),
{\it Twistor Geometry and Non-linear Systems},
Lecture Notes in Mathematics  vol. 970
(Springer-Verlag, 1982).

\item{}
Jones, P.E., and Tod, K.P.,
Class. Quantum Grav. 2 (1985), 565-577.

\item{}
Ward, R.S.,
Class. Quantum Grav. 7 (1990). L95-L98.

\item{[28]} 
Flaherty, E.J.,
Gen. Rel. Grav. 9 (11) (1978), 961-978.

\item{[29]} 
Park, Q-Han,
Integrable deformations of self-dual gravity,
Cambridge preprint DAMTP R-91/5 (July, 1991).

\item{[30]} 
Takasaki, K.,
Volume-preserving diffeomorphisms in
integrable deformations of selfdual gravity,
Kyoto preprint KUCP-0046/92 (March, 1992).

\item{[31]} 
Polchinski, J.,
Nucl. Phys. B362 (1991), 125-140.

\item{}
Avan, J., and Jevicki, A.,
Phys. Lett. 266B (1991), 35-41.

\item{}
Das, S.R., Dhar, A., Mandal, G., and Wadia, S.R.,
Gauge theory formulation of the $c=1$ matrix model:
symmetries and discrete states,
ETH-TH-91/30, IASSNS-HEP-91/52, TIFR-TH-91/44
(September, 1991);
$W$-infinity Ward identities and correlation functions
in the $c=1$ matrix model,
IASSNS-HEP-91/79, TIFR-TH-91/57 (February, 1992).

\item{}
Witten, E.,
Ground ring of two dimensional string theory,
IASSNS-HEP-91/51 (August, 1991).

\item{[32]} 
Hoppe, J., Olshanetsky, M., and Theisen, S.,
Dynamical systems on quantum tori algebras,
Kahlsruhe preprint KA-THEP-10/91 (October, 1991).

\item{}
Hoppe, J.,
Dynamical systems on $W_\infty$ and universal enveloping algebras,
Kahlsruhe preprint KA-THEP-11/91 (December, 1991).

\item{[33]} 
Fairlie, D.B., and Zachos, C.K.,
Phys. Lett. 224B (1989), 101-107.

\item{[34]} 
Hoppe, J.,
Phys. Lett. 250B (1990), 44-48.

\item{[35]} 
Degasperis, A. Lebedev, D., Olshanetskii, M.,
Pakuliak, S., Perelomov, A., Santini, P.,
Recent development for integrable integrodifferential equations,
BONN-HE-90-13 (Nov 1990);
Nonlocal integrable partners to generalized mKdV and
two-dimensional Toda lattice equations in the formalism
of a dressing method with quantized spectral parameter,
BONN-HE-90-14 (Nov 1990).

\item{[36]} 
Lebedev, D., Orlov, A., Pakuliak, S., Zabrodin, A.,
Nonlocal integrable equations as reduction of the Toda hierarchy,
BONN-HE-91-05 (Feb 1991).

\item{[37]} 
Bakas, I., and Kiritsis, E.,
Grassmannian coset models and unitary representation of $W_\infty$,
UBC-PTH-90/16, UMD-PP90-215 (April, 1990).

\item{[38]} 
Witten, E.,
Surprises with topological field theories,
IASSNS-HEP-90/37 (April, 1990).

\bye